\documentclass[pra,twocolumn,floatfix,a4paper,superscriptaddress]{revtex4-2}
\usepackage{bm,color,graphicx,amsmath,txfonts}

\usepackage[colorlinks, citecolor=blue,linkcolor=blue]{hyperref}

\newcommand{\ic}{{i}}
\newcommand{\e}{{e}}

\begin{document}

\title{Coherent feedback control of quantum correlations in cavity magnomechanical system with magnon squeezing}

\author{M. Amazioug}
\affiliation{LPTHE-Department of Physics, Faculty of sciences, Ibn Zohr University, Agadir, Morocco}

\author{S. K. Singh}
\affiliation{Graphene and Advanced 2D Materials Research Group (GAMRG), School of Engineering and Technology,
Sunway University, Selangor Malaysia}
\author{B. Teklu}
\affiliation{Department of Applied Mathematics and Sciences, Khalifa University, Abu Dhabi 127788, UAE}
\affiliation{Center for Cyber-Physical Systems (C2PS), Khalifa University, 127788 Abu Dhabi, UAE}

\author{M. Asjad}
\affiliation{Department of Applied Mathematics and Sciences, Khalifa University, Abu Dhabi 127788, UAE}

\begin{abstract}
 We address a scheme to enhance the quantum correlations in cavity opto-magnomechanical system by using the coherent feedback loop in the presence of magnon squeezing. The proposed coherent feedback-control allows a significant enhancement of the entanglement of three bipartite subsystems, i.e., photon-phonon, photon-magnon and phonon-magnon. We also study the Einstein-Podolsky-Rosen steering and one-way steering in the presence of thermal effects without imposing additional conditions of asymmetric losses or noises in the subsystems. Furthermore, we investigate the sensitiveness of the scheme to the magnon squeezing, and its performance in non-ideal situations in which losses and noises are taken into account.

\end{abstract}

\date{\today}

\maketitle

\section{Introduction} 

Entanglement and Einstein-Podolsky-Rosen (EPR) steering are two quantum resources, which play a crucial role in quantum information processing and communication. Quantum entanglement plays an important role in various applications in quantum information processing, such as quantum teleportation \cite{CHBennett1993}, superdense coding \cite{CHBennett1992}, telecloning \cite{VScarani2005} and quantum cryptography \cite{AKEkert1991}. Many schemes have been proposed over the past decades for processing quantum information such as spins \cite{Michalakis, Szabo}, ions\cite{Retzker, Li, Semiao, Nicacio}, atoms \cite{Raimond,Vollbrecht, jie, asjad1a, Jo, asjad2a}, photons\cite{ Eisenberg, asjad11, Yin, asjad22, Salart, asjad1, Wang, asjad2, berihu22}, phonons \cite{Akram,berihu18, asjad3}. Besides, quantum steering is a class of asymmetric quantum correlations stronger than entanglement \cite{RFWerner1989} but weaker than the violation of Bell's inequality \cite{MTQuitino2015}. The concept of quantum steering was introduced first by Schr\"odinger in the context of the EPR parado \cite{AEinstein1935,ESchrodinger1935} and it can be asymmetric (one-way), and symmetric (two-way) \cite{IKogias2015}.  Steering is then a natural resource for one sided device-independent quantum key distribution \cite{CBranciard2012,NWalk2016}.

In recent years, magnons, as the quanta of collective spin excitations in yttrium iron garnet $(Y_3Fe_5O_{12}, YIG)$, are of paramount importance role due to their high spin density, low damping rate and great tunability. Therefore, cavity magnomechanics has attracted considerable attention and offers a robust platform where ferrimagnetic cristal (e.g., yttrium iron garnet $(YIG)$ sphere) is coupled with a microwave cavity \cite{DDLachanceQuirion2019, HYYuanArxiv}.  In the cavity magnomechanics, a magnon mode (spin wave) is combined with a vibratory deformation mode of a ferromagnet (or ferrimagnet) by the magnetostrictive force, and a microwave cavity mode by the interaction of magnetic dipoles. The magnetostrictive interaction is a dispersive interaction similar to a radiation pressure for a large ferromagnet, where the frequency of the mechanical mode is very lower than the magnon frequency \cite{XZhang2016, ZYFan2022}. 

In this paper, we consider coherent feedback technique \cite{alfred, mat}  to enhance the entanglement and steerability in an opto-magnomechanical system consisting of a cavity contaning $(YIG)$ sphere with the magnon self-Kerr nonlinearity as shown in Fig. \ref{fig1}. We find a significant enhancement of quantum correlations via magnon squeezing which is generated by using the magnon self-Kerr nonlinearity ~\cite{MJCollet1985,SRebic2009}. The magnon self-Kerr nonlinearity [\cite{ass}] can be generated via coupling the magnon mode to a superconducting qubit~\cite{DLachancsQuirion2017}.  We consider the logarithmic negativity \cite{GVidal2002, GAdesso2004} to quantify the quantum entanglement of three bipartite subsystems.
The steerability of the subsystem $A$ by the first subsystem $B$ is used to quantify how much the two entangled bipartite states are steerable. We discuss the enhancement of nonclassical correlations via coherent feedback technique in the presence of the magnon self-Kerr nonlinearity. We show the role of the feedback technique in the presence of the magnon self-Kerr nonlinearity and when $\beta=\pi$ to make the nonclassical correlations very robust to the thermal effects.

The paper is organized as follows. In Sec. \ref{Model}, we give the explicit expression of the Hamiltonian and the corresponding nonlinear quantum Langevin equations of the system. In Sec. \ref{sQLEs}, we provide the linearized quantum Langevin equations for the system. We present a method in Sec. \ref{enta} to quantify entanglement for two-mode continuous-variable (CV) and Gaussian quantum steering. The results and discussions are given in Sec. \ref{res}. Concluding remarks are given in Sec. \ref{conc}

\section{Model}\label{Model}

\begin{figure}[t]
\centering
\includegraphics[width=.45\textwidth]{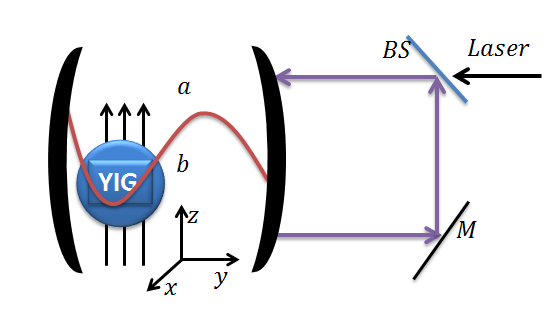} 
\caption{Schematic diagram of a single-mode cavity with feedback loop and a $(YIG)$ sphere with magnon self-Kerr nonliearity. The magnons are embodied by a collective motion of a large number of spins in a macroscopic ferrimagnet, and the magnon mode is directly driven by a microwave source (not shown) to enhance the magnomechanical coupling. The cavity is also driven by an electromagnetic field with amplitude $\Omega$.
The photons and magnons of the cavity are coupled by dipole magnetic interaction, and the magnons and phonons are coupled by magnetostrictive interaction. A microwave field (not shown) is implemented to improve magnon-phonon coupling. At the sphere $(YIG)$, the magnetic field (along the x-axis) of the cavity mode, the driving magnetic field (in the y-direction) and bias magnetic field (z-direction) are common perpendicular.
An input laser light field enters in the cavity across an asymmetric beam splitter (BS). The output field is fully reflected on the $M$ mirror and some of the output field is sent to the cavity by the beam splitter.}
\label{fig1}
\end{figure}

We consider a cavity magnomechanics driven by single coherent laser source and a microwave cavity with coherent feedback as depicted in Fig. \ref{fig1}.
where a yttrium iron garnet (YIG) sphere with the diameter $250-\mu$ m-diameter (Ref.~\cite{XZhang2016}) is placed inside the cavity. In this system, the coupling between magnons and cavity photons is due to the magnetic dipole interaction. The magnetostrictive interaction mediates the coupling between magnons and phonons. The variable magnetisation induced by the magnon excitation within the $(YIG)$ sphere causes the deformation of its geometric structure, which forms the vibrational modes of the sphere, and vice versa~\cite{CKittel1958}. We consider the influence of radiation pressure to be insignificant because the size of the sphere is much smaller than the microwave wavelength. The Hamiltonian of the system is described by the form (with $\hbar=1$)
\begin{eqnarray}\label{Hamiltonian}
\mathcal{H} &=& \omega_a a^{\dag} a + \omega_b b^{\dag} b + \frac{\omega_m}{2} (x^2 + y^2) + \xi (b^\dagger b)^2 + g_{Gb} b^{\dag} b x  \nonumber\\
&+& g_{Ga} (a + a^{\dag}) (b + b^{\dag})  + i \Omega (b^{\dag} e^{-i \omega_0 t}  - b e^{i \omega_0 t} ) \nonumber\\
&+& i \mathcal{E} (a^{\dag} e^{-i \omega_l t}  - a e^{i \omega_l t} ),
\end{eqnarray}
where $a$ ($a^{\dag}$) and $b$ ($b^{\dag}$) ($[O, O^{\dag}]\,{=}\,1$, $O\,{=}\, a, b$) are the annihilation (creation) operators of the cavity and magnon modes, respectively, $x$ and $y$ ($[x, y]\,{=}\,i$) are the dimensionless position and momentum quadratures of the mechanical mode, and $\omega_a$, $\omega_b$, and $ \omega_m$ are respectively the resonance frequency of the cavity, magnon and mechanical modes. $\xi$ is the self-Kerr coefficient. The magnon frequency is determined by the external bias magnetic field $H$ and the gyromagnetic ratio $\kappa$, i.e., $\omega_b=\kappa H$. The single-magnon magnomechanical coupling rate $g_{Gb}$ is small, but the magnomechanical interaction can be improved via driving the magnon mode with a strong microwave field (directly driving the $(YIG)$ sphere with a microwave source \cite{YPWang2018,YPWang2016}).  The coupling rate $g_{Ga}$ between the magnon and microwave can be larger than the dissipation rates $\gamma_a$ and $\gamma_b$ of the cavity and magnon modes respectively, entering into the strong coupling regime, $g_{Ga} > \gamma_a, \gamma_b$. In the frame rotating at the drive frequency $\omega_0$ and applying the rotating-wave approximation (RWA) of the system, $g_{Ga} (a + a^{\dag}) (b + b^{\dag}) \to g_{Ga} (a b^{\dag} + a^{\dag} b)$ (valid when $\omega_a, \omega_b \gg g_{Ga}, \gamma_a, \gamma_G$, which is easily satisfied~\cite{XZhang2016}). The parameter $\Omega =\frac{\sqrt{5}}{4} \kappa \! \sqrt{N} B_0$ represents the Rabi frequency ~\cite{JLi2018} which describes the coupling strength of the drive magnetic field (with $B_0$ and $\omega_0$ are respectively the amplitude and frequency ) with the magnon mode, where $\kappa/2\pi= 28$ GHz/T, and the total number of spins $N=\rho V$ with $V$ the volume of the sphere and $\rho=4.22 \times 10^{27}$ m$^{-3}$ the spin density of the $(YIG)$. The Rabi frequency $\Omega$ is derived under the assumption of the low-lying excitations, $\langle b^{\dag} b \rangle \ll 2Ns$, with $s=\frac{5}{2}$ is the spin number of the ground state Fe$^{3+}$ ion in $(YIG)$.  Then the full dynamics in the presence of coherent feedback and noises  is described by the corresponding quantum Langevin equations (QLEs) 
\begin{eqnarray} \label{QLEs}
\dot{a}&=& - (i \Delta_{fb} + \gamma_{fb}) a - i g_{Ga} b  - \psi \mathcal{E}  + (2 \gamma_a)^{\frac{1}{2}} a_{fb}^{\rm in}, \nonumber \\
\dot{b}&=& - (i \Delta_b + \gamma_b) b - i g_{Ga} a - i g_{Gb} b x -2i \xi b^{\dag}b b + \Omega  + \sqrt{2 \gamma_G} b^{\rm in},\nonumber  \\
\dot{x}&=& \omega_m y, \nonumber  \\
\dot{y}&=& - \omega_m x - \gamma_m y - g_{Gb} b^{\dag}b + \phi, 
\end{eqnarray}
where $\Delta_b=\omega_b-\omega_0$, $\gamma_b$ is the dissipation rate of the magnon mode, $\gamma_m$ is the mechanical damping rate,  $\gamma_{fb}=\gamma_a(1-2\tau\cos{\beta})$ is the modified cavity decay rate and $\Delta_{fb}=\Delta_{a}+2\gamma_a\tau\sin{\beta}$ is the effective detuning with $\Delta_{a}=\omega_{a}-\omega_0$ are respectively the effective cavity decay rate and the detuning with
Here, the quantities $\psi$, $\tau$ denote the transmission and reflection coefficients respectively 
and $\beta$ describes the phase shift generated by the reflectivity of the output field on the mirrors \cite{MAmaziougPLA2020}. The operator $a^{in}_{fb}$ describes the effective input noise operator in the presence of coherent feedback and corresponding description is based on input-output theory \cite{DFWalls1998}. Specifically it can be written as $a^{in}_{fb} =\tau\e^{\ic\beta}a^{out} +\psi a^{in}$, where $a^{in}$ is the input noise operator associatetd with microwave mode with only non-zero correlations $\langle a^{in}(t) a^{in^\dagger}(t')\rangle= n_a(\omega_a)\delta(t-t')$  and $\langle a^{in^\dagger} (t) a^{in}(t')\rangle= (n_a (\omega_a) +1)\delta(t-t')$. Then the corrosponding correlation functions for the effective input noise operator $a^{in}_{fb}$ for the microwave mode can be written as \cite{mamaziougFB2022}
\begin{eqnarray}
\langle a_{fb}^{\rm in}(t) \, a_{fb}^{\rm in \dag}(t')\rangle &=& \psi^2 |1-\tau\e^{\ic\beta}|^2[n_a(\omega_a){+}1] \,\delta(t{-}t'),\nonumber \\
\langle a_{fb}^{\rm in \dag}(t) \, a_{fb}^{\rm in}(t')\rangle &=& \psi^2 |1-\tau\e^{\ic\beta}|^2 n_a(\omega_a) \, \delta(t{-}t').
\end{eqnarray}
Morover, $b^{\rm in}$ and $\phi$ are input noise operators for the magnon and mechanical modes, respectively, which are zero mean and characterized by the following correlation functions~\cite{CWGardiner2000}
\begin{eqnarray}
\langle b^{\rm in}(t) \, b^{\rm in \dag}(t')\rangle &=& [n_b(\omega_b)+1] \, \delta(t{-}t'),\\
\langle b^{\rm in \dag}(t) \, b^{\rm in}(t')\rangle &=& n_b(\omega_b)\, \delta(t{-}t'),\\
\langle \phi(t)\phi(t')\,{+}\,\phi(t') \phi(t) \rangle/2 \,\, &{\simeq}& \,\, \gamma_b [2 n_m(\omega_m) {+}1] \delta(t{-}t').
\end{eqnarray}
The mechanical quality factor ${\cal Q} = \omega_m/\gamma_m \,\, {\gg}\, 1$ is large for a Markovian approximation \cite{Markovian}, where $n_j(\omega_j){=}\big[ {\rm exp}\big( \frac{\hbar \omega_j}{k_B T} \big) {-}1 \big]^{-1} $ $(j{=}a,b,m)$ are the equilibrium mean thermal photon, magnon, and phonon number, respectively. 
\begin{figure*}[!htb] 
\includegraphics[width=1\textwidth]{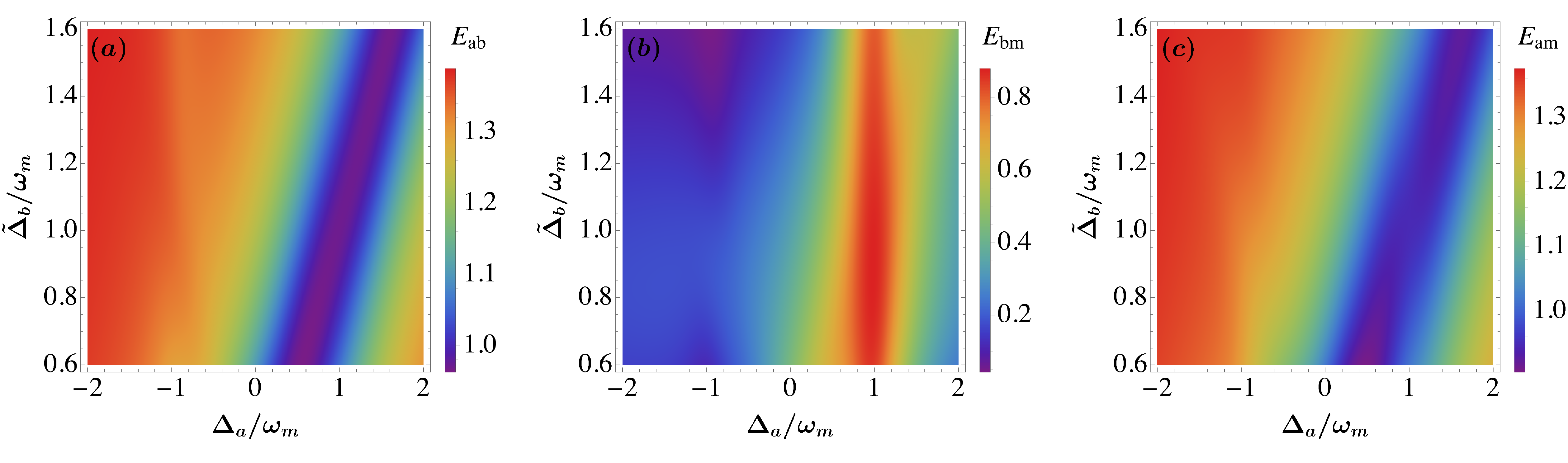}
\caption{(a) Density plot of bipartite entanglement between photon and magnon modes$E_{ab}$, (b) magnon and phonon modes $E_{bm}$ and (c) cavity and phonon modes $E_{am}$ as function of normalize detunings $\Delta_a/\omega_m$ and $\tilde\Delta_b/\omega_m$ for $\tau=0.9$, $T=10$ mK, $\beta=\pi$ and $\xi=\gamma_a$. See text for the other parameters.}
\label{EomM}
\end{figure*}
\section{Linearization of quantum Langevin equations}\label{sQLEs}
Heisenberg-Langevin in Eq.({\ref{QLEs}}) are non-linear in nature and generally cannot be solved analytically. To solve analytical these equations, we use the following linearization scheme. We re-write each operator as a sum of the stationary state mean and a fluctuating quantum operator $O=O_s +\delta O$ ($O\, {=}\, a,b,x,y$), and neglecting second order fluctuation terms when the magnon mode is strongly driven (large amplitude $|\langle b \rangle| \gg 1$ at the steady state), and the cavity field also has a large amplitude $|\langle a \rangle| \gg 1$ via the cavity-magnon beamsplitter interaction. This gives the steady-state solutions according to
\begin{eqnarray} \label{eq5}
\langle b \rangle &=&  \frac{\Omega -ig_{Ga}\langle a \rangle }{ i \tilde{\Delta}_b + \gamma_b },\\ 
\langle a \rangle &=&-\frac{ig_{Ga}\langle b \rangle+i\psi \mathcal{E}}{i\Delta_{fb}+\gamma_{fb}}
\end{eqnarray} 
and for $|\tilde{\Delta}_b|, |\Delta_{fb}| \gg  \gamma_{fb}, \gamma_b$, one gets
\begin{equation}\label{avM}
\langle b \rangle \simeq  \frac{ i \Omega  \Delta_{fb} -i\psi  \mathcal{E}} {g_{Ga}^2  -  \tilde{\Delta}_b \Delta_{fb} },
\end{equation}
where $ \tilde{\Delta}_b = \Delta_b + g_{Gb} \langle x \rangle+ 2i\xi |\langle b \rangle|^2$ is the effective magnon-drive detuning including the frequency shift due to the magnomechanical interaction, and $\tilde{G}_{Gb} = i \sqrt{2} g_{Gb} \langle b \rangle$ is the effective magnomechanical coupling rate, where $ \langle x \rangle = - \frac{g_{Gb}}{\omega_m} \langle b \rangle^2$. The linearized QLEs describing the quadrature fluctuations $\delta X_a=(\delta a + \delta a^{\dag})/\sqrt{2}, \quad \delta Y_a=i(\delta a^{\dag} - \delta a)/\sqrt{2}, \delta X_b=(\delta b + \delta b^{\dag})/\sqrt{2}, \delta Y_b=i(\delta b^{\dag} - \delta b)/\sqrt{2}, \delta x$ and $\delta y$ can be written in compact matrix form as
\begin{equation}
\dot{u} (t) = \mathcal{L} u(t) + \mu (t) ,
\end{equation}
where $u(t)=\big[\delta X_a (t), \delta Y_a (t), \delta X_b (t), \delta Y_b (t), \delta x (t), \delta y (t)\big]^T$ is vector of quadrature fluctuation operators, $\mu (t) = \big[ \!\sqrt{2\gamma_a} X_a^{\rm in} (t), \sqrt{2\gamma_a} Y_a^{\rm in} (t), \sqrt{2\gamma_b} X_b^{\rm in} (t), \sqrt{2\gamma_b} Y_b^{\rm in} (t), 0, \phi(t) \big]^T$ is the vector of input noise operators, and the drift matrix $\mathcal{L}$ can be written as
\begin{equation}\label{AAA}
\mathcal{L} =
\begin{pmatrix}
-\gamma_{fb}  &  \Delta_{fb}  &  0 &  g_{Ga}  &  0  &  0  \\
-\Delta_{fb}  & -\gamma_{fb}  & -g_{Ga}  & 0  &  0  &  0  \\
0 & g_{Ga}  & -\gamma_b+\xi  & \tilde{\Delta}_b & -\tilde{G}_{Gb} & 0 \\
-g_{Ga}  & 0 & -\tilde{\Delta}_{b} & -\gamma_b-\xi &  0  &  0  \\
0 &  0  &  0  &  0  &  0  &  \omega_m  \\
0 &  0  &  0  &  \tilde{G}_{Gb}  & -\omega_m & -\gamma_m  \\
\end{pmatrix}.
\end{equation}
The drift matrix in Eq.~\eqref{AAA} is provided under the condition $|\tilde{\Delta}_b|, |\Delta_{fb}| \gg  \gamma_{fb}, \gamma_b$. In fact, we will show later that $| \tilde{\Delta}_b|, |\Delta_{fb}| \simeq \omega_m  \gg  \gamma_{fb}, \gamma_b$ [see Fig.~\ref{fig1} (b)] are optimal for the presence of all bipartite entanglements of the system. Note that Eq.~\eqref{eq5} is intrinsically nonlinear since $ \tilde{\Delta}_b$ contains $ |\langle b \rangle|^2 $. However, for a given value of $\tilde{\Delta}_b$ (one can always tune $\tilde{\Delta}_b$ by adjusting the bias magnetic field) $\langle b \rangle$, and hence $\tilde{G}_{Gb}$, can be achieved straightforwardly.
\begin{figure*}[!htb]
\includegraphics[width=1\textwidth]{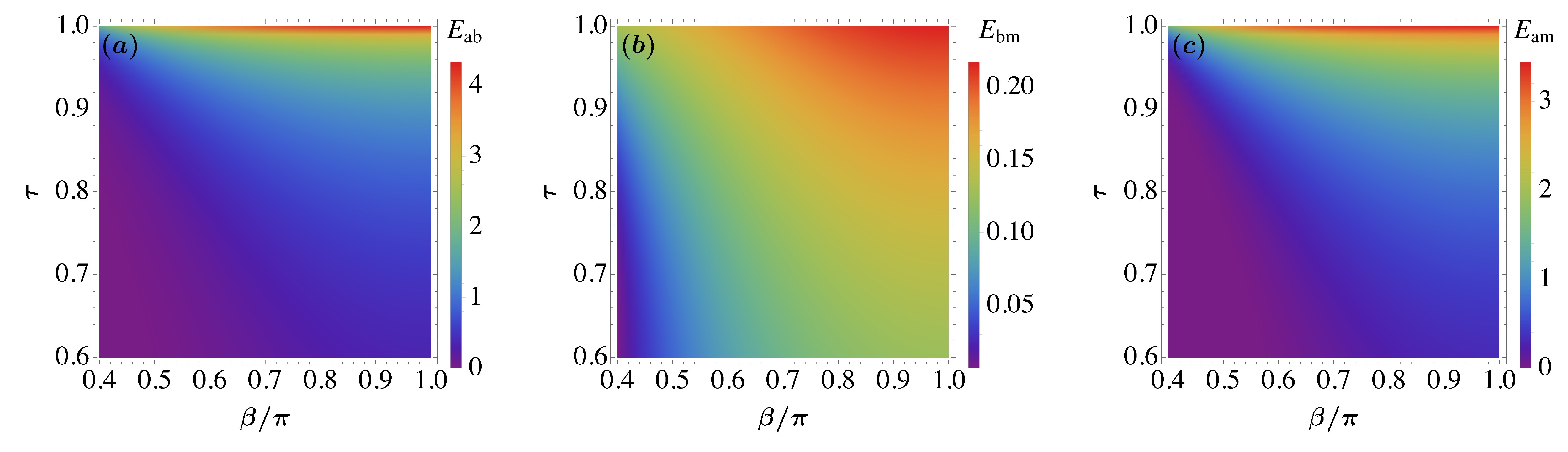}
\caption{(a) Density plot of bipartite entanglement between photon and magnon modes $E_{ab}$, (b) magnon and phonon modes $E_{bm}$ and (c) cavity and phonon modes $E_{am}$ versus the reflectivity parameter $\tau$ and phase $\beta$ for  $\tilde\Delta_b = 0.9 \omega_m$, $\Delta_a = - \omega_m$ and $\xi=\gamma_a$. See text for the other parameters.}
\label{EaT}
\end{figure*}
\section{Entanglement and steerabilities}\label{enta}
\begin{figure*}[!tb]
\centering
\includegraphics[width=1\textwidth]{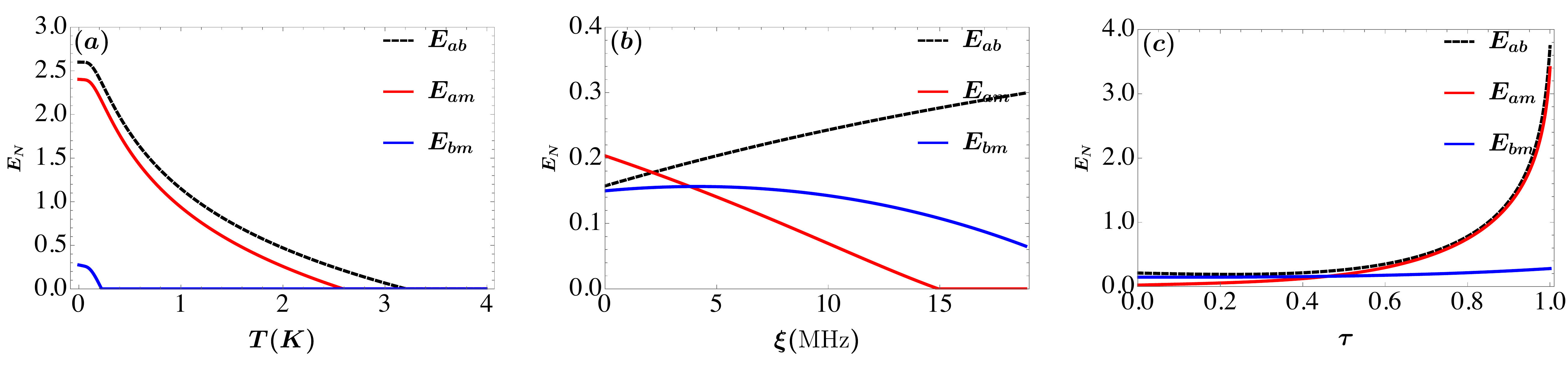}
\caption{(a) Plot of photon and magnon modes ($E_{ab}$), cavity and phonon modes ($E_{am}$) and magnon and phonon modes ($E_{bm}$) as a function of temperature $T$, self-Kerr coefficient $\xi$ (see the Fig. (b)) and reflectivity parameter $\tau$ (see the Fig. (c)). We take $G_{Gb}/2\pi = 4.8$ MHz, $\Delta_a = - \omega_m$ and $\tilde\Delta_b = 0.9 \omega_m$. The reflectivity parameter is $\tau=0.98$ and $\tau=0.4$ in Fig. (a) and Fig. (b) respectively. In Figs. (b)-(c) the temperature is $T=10$ mK and in Figs. (a)-(c) the magnon self-Kerr nonlinearity coefficient is $\xi=\gamma_a$. See text for the details of the other parameters.}
\label{ETxiTau}
\end{figure*}
The steady state evolution of the quantum fluctuations of the system is a continuous variable (CV) three-mode Gaussian state is completely characterized by a $6 \times 6$ covariance matrix (CM) $\mathcal{V}$, where $\mathcal{V}_{ij}=\frac{1}{2}\langle u_i(t) u_j(t') + u_j(t') u_i(t) \rangle$ ($i,j=1,2,...,6$) of the covariance matrix (CM) $\Delta$ satisfies \cite{DVitali2007, PCParks1993}
\begin{equation}\label{Lyap}
\mathcal{L} \mathcal{V}(t) + \mathcal{V}(t) \mathcal{L}^T =  - \mathcal{K},
\end{equation}
where $\mathcal{K}={\rm diag} \big[\gamma_a \psi^2 |1-\tau e^{i\beta}|^2 (2n_a+1), \gamma_a\psi^2 |1-\tau e^{i\beta}|^2 (2n_a+1), \gamma_b (2n_b+1),  \gamma_b (2n_b+1), 0,  \gamma_m (2n_m +1 ) \big]$ is the diffusion matrix, which is defined through $\langle  \mu_i(t) \mu_j(t') + \mu_j(t') \mu_i(t) \rangle/2 = \mathcal{K}_{ij} \delta (t-t')$.
The covariance matrix $\sigma_{AB}$ of two modes $A$ and $B$ may be written as
\begin{equation} \label{eq:Sigmamm}
\sigma_{AB}=
\begin{pmatrix}
	\mathcal{A} & \mathcal{C}  \\
    \mathcal{C}^T & \mathcal{B}  
\end{pmatrix}.
\end{equation} 
The $2\times 2$ sub-matrices $\mathcal{A}$ and $\mathcal{B}$ in Eq. (\ref{eq:Sigmamm}) describe the autocorrelations of the two modes and $2\times 2$ sub-matrix $\mathcal{C}$ in Eq. (\ref{eq:Sigmamm}) denotes the cross-correlations of the two modes. Characterizing, quantifying and classifying quantum correlations in multipartite quantum systems are one of the most problematic issues in quantum information and especially in optomagnomechanical systems when information is encoded in continuous variables (CV). The entanglement in CV system can be quantified by 
 by using the logrithmic negativity  $E_N$~\cite{GVidal2002,GAdesso2004}
\begin{equation} 
	E_N = \max[0,-\log(2\Lambda^-)],
\end{equation}
where $\Lambda^-= \sqrt{\mathcal{X}-(\mathcal{X}^2-4\det\sigma)^{1/2}}/\sqrt{2}$ being the minimum symplectic eigenvalue of partially transposed covariance matrix of two mode Gaussian states, with $\mathcal{X}=\det \mathcal{A}+\det \mathcal{B}-\det \mathcal{C}$. The two subsystems are entangled if $E_N>0$. For the two-mode Gaussian state (\ref{eq:Sigmamm}), the system is separable if $\Lambda^-<1/2$, where $\Lambda^-$ being the smallest symplectic eigenvalue of partial transposed covariance matrix Eq. (\ref{eq:Sigmamm}). Another quantum correlation quantifier which  is of paramount importance in optomagnomechanical systems is the quantum steering. The steerability of Bob $(B)$ by Alice $(A)$ $(A \to B)$ for a $(n_A + n_B)$ mode Gaussian state can be quantified by \cite{IKogias2015}
\begin{equation} 
	S^{A\to B} (\sigma_{AB}) = \max\left[0,-\sum_{j:\bar \nu_j^{AB/ A}<1}ln\left(\bar\nu_j^{AB/ A}\right)\right],
\end{equation}
where $\bar \nu_j^{AB/ A}$ $( j = 1, . . . , m_B)$ are the symplectic eigenvalues of $\bar \sigma_{AB/ A} = B - C^T A^{-1} C$, derived from the Schur complement of $A$ in the covariance matrix $\sigma_{AB}$. The steerability of Alice by Bob $[S^{A\to B} (\sigma_{AB})]$ can be obtained by swapping the roles of $A$ and $B$. We notice that a non-separable state is not always steerable but a steerable state is always non separable state. Thus we have two possibilities between $A$ and $B$ : ($i$) if $\mathcal{S}^{A \to B}=\mathcal{S}^{B \to A}=0$ Alice can't steer Bob and vice versa even if they are entangled (i.e. no-way steering), ($ii$) if $\mathcal{S}^{A \to B}>0$ and $\mathcal{S}^{B \to A}=0$ or $\mathcal{S}^{A \to B}=0$ and $\mathcal{S}^{B \to A}>0$ as one-way steering, i.e. Alice can steer Bob but Bob can't steer Alice and vice versa, and ($iii$) if $\mathcal{S}^{A \to B}=\mathcal{S}^{B \to A}>0$ Alice can steer Bob and vice versa (i.e. two-way steering). In addition, the measurement of Gaussian Steering is always bounded by the entanglement. In order to check the asymmetric steerability of the two mode Gaussian state, we introduce the steering asymmetry which is defined as
\begin{equation} 
	S(AB)=|S^{A\to B} - S^{B\to A}|.
\end{equation}
\section{Results and Discusion} \label{res} 
In this section, we show the results and discuss the evolution of quantum correlations of the system by considering experimentally accessible parameters reported in \cite{mamaziougFB2022, JLi2018}: $\omega_a/2\pi= 10$ GHz, $\omega_m/2\pi= 10$ MHz , $\gamma_m/2\pi = 100$ Hz, $\gamma_a/2\pi = \gamma_b/2\pi = 1$ MHz, $g_{Ga}/2\pi= G_{Gb}/2\pi=3.2$ MHz, and at low temperature $T = 10$ mK. $G_{Gb} = 2\pi \times 3.2$ MHz implies the drive magnetic field $B_0 \approx 3.9\times 10^{-5}$ T for $g_{Ga}\approx 2\pi \times 0.2$ Hz, corresponding to the drive power $P =8.9$ mW.
\begin{figure*}[htb]
\includegraphics[width=1\textwidth]{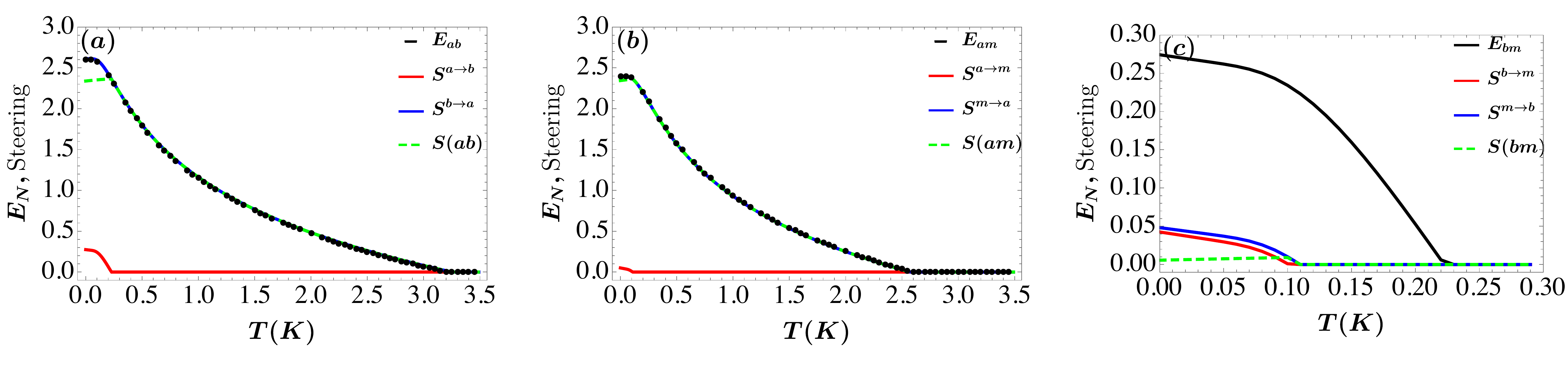}
\caption{Plot of bipartite entanglement, Gaussian quantum steering and  asymmetric quantum steering between photon and magnon modes (a) $E_{ab}$, $S^{a\to b}$ and $S^{b\to a}$ and $S(ab)$, cavity and phonon modes (b) $E_{am}$, $S^{a\to m}$ and $S^{m\to a}$ and $S(am)$ and magnon and phonon modes (c) $E_{bm}$, $S^{b\to m}$ and $S^{m\to b}$ and $S(bm)$, Gaussian quantum steering between photon and magnon modes ($S^{a\to b}$ and $S^{b\to a}$) as a function of the temperature $T$ . We take $G_{Gb}/2\pi = 4.8$ MHz, $\beta=\pi$, $\tau=0.98$, $\xi=\gamma_a$, $\tilde\Delta_b = 0.9 \omega_m$ and $\Delta_a = - \omega_m$. See text for the details of the other parameters.}
\label{ESAS}
\end{figure*}

We present in Fig. (\ref{EomM}), the steady state of the three bipartite entanglement $E_{ab}$ (between the cavity and magnon mode), $E_{bm}$ (between the magnon and mechanical mode) and $E_{am}$ (between the cavity and mechanical mode) versus the detunings $\Delta_a$ and $\tilde\Delta_b$ in the presence of coherent feedback loop with the magnon self-Kerr nonlinearity. We observe, that the entanglement is very strong ($E_{ab}>1.3$, $E_{bm}>0.8$ and $E_{am}>1.3$) in comprising with the results in Ref. \cite{mamaziougFB2022,JLi2018}. The maximum value of entanglement of the three bipartite is improves via coherent feedback loop and the magnon self-Kerr nonlinearity when $\beta=\pi$. We remark, when $\Delta_a = - \omega_m$ and $\tilde\Delta_b=0.9\omega_m$ the entanglement $E_{ab}$ and $E_{am}$ are maximum while $E_{bm}\approx 0.2$. In Fig. (\ref{EaT}) we plot there bipartite entanglements $E_{ab}$, $E_{bm}$ and $E_{am}$ as a function of the reflectivity $\tau$ and $\beta$ as shown  We remark that the entanglement is increasing with $\tau$ and $\beta$ and it robust when $\beta=\pi$. Moreover, the entanglement is achieved its maximum value when $\gamma_{fb}=\gamma_a(1+2\tau)$.

In Fig. (\ref{ETxiTau}) we plot the entanglements of the three bipartite $E_{ab}$, $E_{am}$ and $E_{bm}$ versus different parameters. We remark that the existing of genuig tripartite entanglement when all bipartite entanglement are non-vanishing as illustrated in Fig. (\ref{ETxiTau}). We notice, the entanglement is robust against temperature as depicted in Fig. (\ref{ETxiTau})(a) and survive above 3K. We observe that the entanglement of all the subsystem is diminishes due to decoherence phenomenon \cite{decoherence}. Moreover, the entanglement between photon-magnon and photon-phonon persists for temperature $T>3$ K and $T\approx 2.5$ K respectively, whereas, the entanglement between magnon-phonon is vanishes at lower temperatures ($T\approx 0.2$ K) even this temperature is the maximum achieved in the Ref. \cite{JLi2018}. One can say that the entanglement between photon-magnon and photon-phonon is stronger than the entanglement between magnon-phonon. The entanglement between photon-magnon and magnon-phonon increases with increasing the magnon self-Kerr nonlinearity coefficient $\xi$, instead the entanglement between photon-phonon decreases as illustrated in Fig. (\ref{ETxiTau})(b). The entanglement $E_{ab}\approx 0.25$ for $\xi=10^7$ Hz in comprising $E_{ab}\approx 0.125$ in comprising with the results in Ref. \cite{JLi2018}. We remark in Fig. (\ref{ETxiTau})(c) the enhancement of all three bipartite entanglement by coherent feedback technique. The maximum value reached by entanglement between photon-magnon and photon-phonon is very important than which obtained in Ref. \cite{mamaziougFB2022,JLi2018}.

In Fig.(\ref{ESAS}), we plot for each bipartite the entanglement, the Gaussian steering $S^{A\to B}$, $S^{B\to A}$ and the asymmetric steering versus the temperature $T$. The entanglement and steerabilities diminish quickly with temperature due to the decoherence phenomenon. We note, the one way quantum steering is more robust than two way quantum steering and it survive for a larger value of temperature $T$. The entangled state is not always steerable state instead steerable state must be entangled i.e. when $S^{A\to B} = S^{B\to A} > 0$ and $E_N > 0$ is the witnesses of existence of Gaussian two-way steering, such that the subsystem of two subsystem are entangled but are steerable only from $A$ to $B$ and from $B$ to $A$ \cite{IKogias2015} and no-way steering appears when $S^{A\to B} = S^{B\to A} = 0$ and $E_N > 0$ as depicted in Fig.(\ref{ESAS})(c). The measurement of Gaussian steering is always bounded by the entanglement $E_N$ as also discussed in \cite{AmaziougOptik18}. Finally, the asymmetric steering $SA$ is always less than $ln(2)$, which is maximal when the state is nonsteerable in one-way i.e. $S^{A\to B}>0$ and $S^{B\to A}=0$ or $S^{A\to B}=0$ and $S^{B\to A}>0$ and it decreases with increasing steerability in either way \cite{IKogias2015}. In Fig.(\ref{ESAS})(a) the steering from the photon mode to the magnon mode $S^{a\to b}$ has a similar behavior to $E_N$ it decreases from its maximum value to zero when $T>3$ K. Besides, one-way steering appears when $T>0.2$ K, i.e. $S^{a\to b}>0$ and $S^{b\to a}=0$ as expected in Fig. (\ref{ESAS})(a). Moreover, the steering from the magnon mode to the photon mode $S^{b\to a}$ is diminishes quickly to remains zero for $T>0.2$ K as depicted in Fig.(\ref{ESAS})(a). Otherwise, when the temperature $T<0.2$ K, the two-way steering occurs between optical mode and the magnon mode, i.e. $S^{a\to b}>0$ and $S^{b\to a}>0$ ($S(ab)=0$).  The steerability between photon mode and the phonon mode is always remains one-way steering, i.e. $S^{a\to m}>0$ ($S^{m\to a}=0$) when $T>0.2$ K as implemented in Fig.(\ref{ESAS})(b). The steerability between the magnon mode and phonon mode is approximately remains two-way steering and $S^{b\to m}>S^{m\to b}$  when $T<0.10$ K and no-way steering ($S^{b\to m}=0$ and $S^{m\to b}=0$ ($S(bm)=0$) when $T>0.10$ K as shown in Fig.(\ref{ESAS})(c).
\section{Conclusions} \label{conc}
 In conclusion we have studied how coherent feedback loop improves the quantum correlations between three bipartite subsystems in the presence of the magnon self-Kerr nonlinearity in cavity magnomechanics systems.  We quantify steerability by using Gaussian quantum steering and show that Gaussian steering remains bounded by entanglement, i.e. the steerable modes are strictly entangled but the entangled modes are not necessarily steerable. We have found one way-steering between photon-magnon and photon-phonon. However, the steerability bewteen magnon-photon is always two-way.  The entanglement and steerabilities are shown robust against the temperature, where entanglement may persist above 3 K in the case of photon-magnon, and approximately equal 2.5 K for photon-phonon. Moreover, The entanglement and steerabilities between magnon-phonon is fragile under thermal effects. Our proposed scheme to improve entanglement can be of interest for various applications in quantum information processing.

\end{document}